\documentclass[prl,twocolumn,showpacs,preprintnumbers,amsmath,amssymb]{revtex4}
\usepackage{graphicx}% Include figure files
\usepackage{dcolumn}% Align table columns on decimal point
\usepackage{bm}% bold math
\usepackage{textcomp}

\begin{document}

\title{Phase coherent splitting of Bose-Einstein condensates with an integrated magnetic grating}

\author{A. G\"unther}
\author{S. Kraft}
\author{C. Zimmermann}
\author{J. Fort\'agh}
\email{fortagh@uni-tuebingen.de}
\affiliation{Physikalisches Institut der Universit\"at T\"ubingen,
Auf der Morgenstelle 14, 72076 T\"ubingen, Germany}

\date{\today}

\begin{abstract}
We report the phase coherent splitting of Bose-Einstein
condensates by means of a phase grating produced near the surface
of a micro-electronic chip. A lattice potential with a period of
$4\mu\text{m}$ is generated by the superposition of static and
oscillating magnetic fields. Precise control of the diffraction is
achieved by controlling the currents in the integrated conductors.
The interference of overlapping diffraction orders is observed
after 8ms of propagation in a harmonic trap and subsequent
ballistic expansion of the atomic ensemble. By analyzing the
interference pattern we show a reproducible phase relation between
the diffraction orders with an uncertainty limited by the resolution
of the diffraction grating.
\end{abstract}

\maketitle

Atom optics and matter wave interferometry have made enormous
progress during the last years. Numerous experiments have been
performed with Bose-Einstein condensates serving as phase coherent
sources of atomic matter waves \cite{Kasevich2002}. As a part of these, diffraction
from standing and moving optical lattice potentials has been demonstrated to preserve the
condensate's coherence properties and is now established as a
versatile tool for coherent manipulation 
\cite{Hagley1999,Stenger1999,Kozuma1999,Simsarian2000,Bongs2001,Torii2000}.
The implementation of similar diffraction scenarios on atomic
micro chips is thus one of the most intriguing challenges of
integrated atom optics. It holds great promise for transportable,
high precision inertial force and gravity sensors
\cite{Kasevich2002}, and for building complex atom optical
circuits, e.g. quantum registers for information processing
\cite{Birkl2007}. Although Bose-Einstein condensates
are routinely loaded into magnetic micro traps
\cite{Fortagh2007}, it is technically demanding to
integrate atom optical elements on a chip that allows for coherent
manipulation of matter waves. Recently, diffraction from optical
\cite{Wang2005} and magnetic lattices \cite{Guenther2005b}, as
well as dynamical splitting of condensates in double well
potentials \cite{Shin2005,Schumm2005,Jo2007} have been studied on micro
chips.

\begin{figure}
\centerline{\scalebox{0.5}{\includegraphics{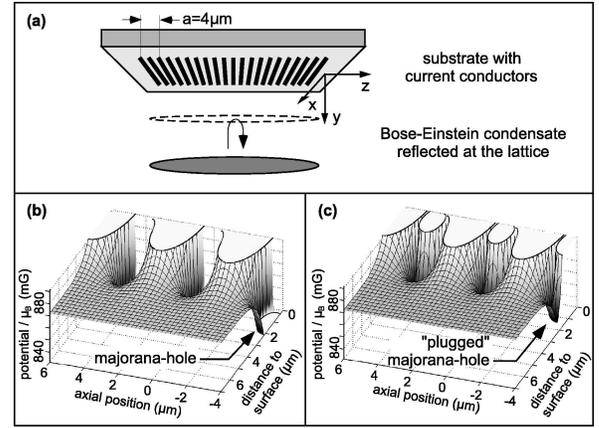}}}
\caption{(a) A condensate is scattered on an integrated magnetic
grating by a controlled center of mass oscillation inside the
magnetic trap. (b) Static magnetic grating: Lattice potential generated by static magnetic fields of
parallel chip conductors carrying opposite currents and the field of the 
magnetic trap. Steep potential wells and hills alternate next to
neighboring conductors. (c) Radio frequency controlled magnetic
grating: Superposition of an oscillating magnetic field parallel
to the lattice conductors generates adiabatic potentials at the
chip surface. The plot reveals wells with limited depth
and repulsive potential walls next to each conductor.}
\label{fig:Figure1}
\end{figure}

In this Letter, we demonstrate phase coherent splitting of Bose-Einstein condensates by means of diffraction on an integrated magnetic grating and realize a novel interferometric scheme. The magnetic grating is generated near the surface of micro fabricated current conductors (Fig.\ref{fig:Figure1}).
We use static and oscillating magnetic fields for shaping and controlling the grating potential. Thereby, sensitive control of the populations of diffraction orders is achieved. Due to the relatively large lattice constant of $4\mu\text{m}$ and a correspondingly small recoil momentum, the
diffraction orders of a condensate remain spatially overlapped during ballistic
expansion. We observe their matter wave interference with
a highly reproducible spatial fringe position. This provides
experimental proof that diffraction from the magnetic grating preserves phase relation
between the diffraction orders of a Bose-Einstein condensate.
Magnetic gratings as described in this article will be important
for phase coherent manipulation of matter waves in integrated atom
optical circuits and for the realization of compact atom interferometers.
In addition, microfabricted gratings extend the range of lattice constants accessible with optical
standing wave fields \cite{Hagley1999}. In particular, the size
and the structure of the unit cell can be defined by the geometry
of chip conductors. 

\begin{figure}
\centerline{\scalebox{0.45}{\includegraphics{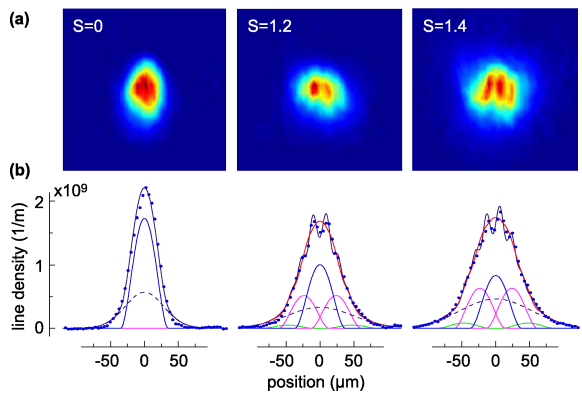}}}
\caption{(Color online) (a) Absorption images of diffracted
condensates taken after 8ms propagation in the magnetic trap
followed by 20ms of ballistic expansion. Diffraction is observed
as a broadening of the cloud and by the emergence of interference
fringes. The fringe contrast is limited by the optical resolution of our imaging system. (b) Vertically integrated absorption profiles of the
condensate (dots). The data is fitted by the incoherent sum of
diffraction orders to obtain $S$ (red line). Illustrated
is also the position and the population of individual diffraction
orders (solid lines) and the thermal background (dashed line). Details of the line density profiles are explained by
fitting the data with a coherent superposition of the diffraction
orders (blue line).} \label{fig:Figure2}
\end{figure}

The experiment is illustrated in Fig.\ref{fig:Figure1}a. An
elongated Bose-Einstein condensates of about $1.2\times 10^5$
$^{87}\text{Rb}$ atoms in the $F\!=\!2$, $m_F\!=\!2$ hyperfine ground
state is prepared in a magnetic micro trap which is placed
$30\mu\text{m}$ below the magnetic grating
\cite{Guenther2005a}. The trap is characterized by the
axial and radial oscillation frequencies of
$(\omega_a,\omega_r)=2\pi\times(16,76)\text{s}^{-1}$, and the
offset field of $B_z=0.87\text{G}$ in the trap center. Diffraction
of the condensate is initiated by suddenly reducing the trap-surface 
separation by a controlled displacement. 
The condensate interacts with the lattice
at the upper turning point of the radial center of mass oscillation inside 
the trap (y-direction) followed by a swing
back towards the trap center. The trap is turned off 12ms after
the displacement and the condensate is detected after 20ms of
ballistic expansion by absorption imaging. The imaging axis is
parallel to the lattice conductors ($x$-direction). Diffraction is
observed as a broadening of the cloud. Additionally, the emergence
of interference fringes is observed (Fig.\ref{fig:Figure2}).

The magnetic lattice potential is generated by a set of parallel
current conductors on the chip surface. The $1\mu\text{m}$ wide
gold conductors are micro fabricated on a silicon substrate with a
separation of $1\mu\text{m}$ between nearest neighbors \cite{Guenther2005a}.
Adjacent conductors carry constant
currents of opposite sign: $\pm 0.2\text{mA}$. The magnetic field
of this wire configuration is superposed onto the trapping field
including the offset field along $z$. The resulting magnetic field
modulus has a periodicity of $4\mu\text{m}$ along $z$ and the
modulation amplitude decreases exponentially from the chip surface
on the length scale of the lattice constant. Steep
potential wells and hills emerge close to the lattice wires
(Fig.\ref{fig:Figure1}b).

\begin{figure}
\centerline{\scalebox{0.42}{\includegraphics{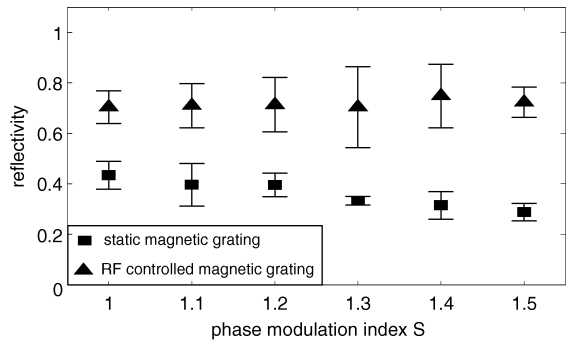}}}
\caption{Reflectivity of the static (squares)
and the radio frequency controlled (triangles) magnetic grating. The
data is plotted against the phase modulation index $S$. The error
bars result from averaging 3-10 individual measurements (standard
deviation). $S$ is fitted with an accuracy of about 0.1.}
\label{fig:Figure3}
\end{figure}

Diffraction on such a magnetic lattice potential
\cite{Guenther2005b} is associated with a significant loss of
atoms (Fig.\ref{fig:Figure3}). The dominant loss mechanism is due
to the steep magnetic field gradients inside the potential wells
of the lattice (Fig.\ref{fig:Figure1}b). Atoms entering the wells
are lost due to Majorana spin flips which result from the sudden
break down of the adiabatic condition along the atom's trajectory
\cite{Sukumar1997}. Majorana loss may be tolerable if the
experiment relies on a single diffraction pulse. Numerous
interferometric schemes, however, require sequences of diffraction
pulses \cite{Simsarian2000,Bongs2001,Torii2000} and the loss of
atoms represents a limitation.

We eliminate losses by introducing radio frequency (rf) controlled
adiabatic potentials
\cite{Schumm2005,Lesanovsky2006}
to ``plug'' the Majorana hole and observe enhanced reflectivity on
the magnetic lattice. An oscillating magnetic field of $10\text{mG}$
amplitude is superposed to the static field of the lattice. The
polarization is parallel to the lattice conductors and the
frequency of 585kHz is below the Larmor frequency of 610kHz
corresponding to the offset field $B_z$ in the trap center. The
resulting adiabatic lattice potential features wells of limited
depth and the initially steep, attractive potential slopes are
converted into repulsive potential walls (Fig.\ref{fig:Figure1}c).
We characterize the reduction of Majorana losses by detecting the
total number of atoms reflected from the grating. The data is
plotted against the phase modulation index $S$
(Fig.\ref{fig:Figure3}). $S$ characterizes the population of the
different diffraction orders as described below; e.g. for $S \approx
1.44$, the $0th$ and $\pm 1st$ diffraction orders have about the
same population. The population of diffraction orders is varied by
the displacement of the trap. The rf field is
activated simultaneously to the displacement and is deactivated
prior to turning off the magnetic trap for ballistic expansion.
The rf controlled magnetic grating displays high
reflectivity compared to its static counterpart. From this result
we deduce that the motion of atoms is adiabatic in the rf controlled lattice potential. The high reflectivity
diffraction grating allows phase coherent manipulation of
Bose-Einstein condensates and the realization of a matter wave
interferometer on a chip. The observation of a deterministic phase relation 
between diffraction orders of the condensate is the central result of this
Letter.

The interaction of the condensate with the magnetic grating is
described by the model of matter wave diffraction in the
Raman-Nath regime \cite{Henkel1994}. In particular, we assume a
short interaction with a one-dimensional grating. 
This way the action of the grating is reduced
to a phase modulation of the condensate wavefunction:
\begin{eqnarray}
\Psi(z) & = & \sqrt{\rho_0(z)}\exp{\left(-\frac{i}{\hbar}\int U(z,t)\text{d}t\right)}\label{Eq:imprint1}\\
            & = & \sqrt{\rho_0(z)}\exp{\left(-iS\cos\left(kz+\phi\right)\right)\label{Eq:imprint2}}.
\end{eqnarray}
Here, $\rho_0$ is the condensate's density, $U(z,t)=u(t)\cdot
\cos{(kz+\phi)}$ is the lattice potential, and the wave vector
$k=2\pi/a$ is given by the lattice constant $a$. $\phi$
characterizes a shift of the lattice position with respect to the
center of the condensate. The amplitude $u(t)$ accounts for the
varying modulation depth during reflection as seen by the atom
along its trajectory. The accumulated phase can be summarized
using the phase modulation index $S$.

The validity of this approximation is verified by numerical
solution of the three-dimensional Gross-Pitaevskii equation \cite{Judd2007}.
It has been shown that the Raman-Nath approximation
%for determining the amplitude of the diffraction orders
holds for $E_r\tau\ll\hbar/S$, where $E_r=(\hbar k)^2/2m$ is the
recoil energy and $\tau$ the effective interaction time with the
lattice \cite{Henkel1994}. With $\tau < 0.2ms$ this condition is
well satisfied in our experiment.

The wavefunction (Eq.\ref{Eq:imprint2}) can be expanded into
momentum eigenfunctions of the axial motion:
\begin{equation}
\Psi(z)=\sqrt{\rho_0(z)}\sum_{n=-\infty}^{\infty} e^{in(\phi-\pi/2)}J_n(S)e^{inkz}\label{Eq:imprint3}.
\end{equation}
Here, $J_n$ are Bessel functions of first kind and $n$ is an
integer. The total wavefunction $\Psi=\sum \Psi_n$ is therefore a
superposition of the discrete momentum eigenfunctions $\Psi_n$,
describing the wavefunction of the $n^{\text{th}}$ diffraction
order. The probability for an atom to be diffracted into the
$n^\text{th}$ order is proportional to $\left|J_n(S)\right|^2$,
and the relative phase between neighboring orders is
$(\phi-\pi/2)$.

\begin{figure}
\centerline{\scalebox{0.45}{\includegraphics{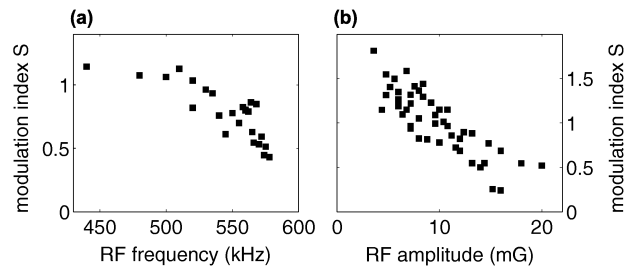}}}
\caption{Control of diffraction by means of the oscillating
magnetic field. The diffraction is initiated by a displacement $d$ of the
trap center towards the grating. (a) At constant amplitude of 10mG the frequency is used to control the diffraction. $d=14\mu\text{m}$. (b) At constant frequency of 585kHz the amplitude is used to control the diffraction. $d=14.5\mu\text{m}$.} \label{fig:Figure4}
\end{figure}

By exploiting the dependence of $S$ on the potential modulation
(Eq.\ref{Eq:imprint1} and \ref{Eq:imprint2}) we control the
population of the diffraction orders with the radio frequency
field. The frequency and the amplitude of the oscillating field
determines the coupling between internal atomic states and with it
the depth and shape of the potential wells. Radio frequency
control of diffraction (Fig.\ref{fig:Figure4}) may be particulary
convenient for integrated atom-optical devices when sequences of
diffraction pulses are required. The phase modulation index can be
electronically changed while the condensate performs a constant
center of mass oscillation in the trap and interacts with the grating in
subsequent oscillation periods.

The experimental results are derived from absorption images as
illustrated in Fig.\ref{fig:Figure2}. For the analysis, the
images are summed up along the y-coordinate, perpendicular to the
course of diffraction. The resulting line density profile is
fitted first by the incoherent superposition of the diffraction
orders, $\rho_{\text{incoh}}\!=\!\sum \left|\Psi_n\right|^2$, to
obtain the phase modulation index $S$. Here, we assume that, at the time of absorption
imaging, the centers of neighboring diffraction orders are
separated by a distance of $\Delta z\!=\!\int\hbar k(t)/m
\text{d}t\!\approx\! 24\mu\text{m}$ due to 8ms propagation in the trap
followed by 20ms ballistic expansion \cite{Note2}. 
The repulsive atomic
mean-field interaction is included by assuming the usual ballistic
expansion of condensates \cite{Castin1996}. Thermal atoms are taken into account by a Gaussian contribution to
the line density profile. The incoherent sum of diffraction orders
fits well to the overall cloud envelope but it does not explain
further details of the cloud.

Unlike diffraction from standard optical lattice potentials
\cite{Hagley1999,Stenger1999,Kozuma1999,Simsarian2000,Bongs2001,Torii2000,Wang2005}, the recoil energy $E_r=(\hbar k)^2/2m$
transferred by the $4\mu\text{m}$ lattice is smaller than the
chemical potential of the condensate. The diffraction orders do
not separate entirely during ballistic expansion and the spatial
distribution of the condensate is given by the coherent
superposition of the diffraction orders \cite{Note3}:
$\rho_{\text{coh}}=\left|\sum\Psi_n\right|^2$. We observe high
contrast interference fringes of overlapping diffraction orders
for phase modulation indices between $1.1-1.6$
(Fig.\ref{fig:Figure2}). The fringes are dominated by the
interference of the $0th$ and $\pm 1st$ orders with populations
that are comparable in this range. 

The phase of the individual diffraction orders after ballistic
expansion results from evaluating the action integral
\cite{Henkel1994}. It contains various contributions: the
parabolic phase profile characterized by a curvature $\alpha$
\cite{Castin1996}, the center of mass momentum of the diffraction
order given by $k$ \cite{Simsarian2000}, and phase contributions
$\phi_{\text{trap}}$ and $\phi_{\text{tof}}$ accumulated during
propagation in the trap and during time of flight. Interference of
overlapping diffraction orders results from the coherent
superposition of the wavefunctions
\begin{eqnarray}
\Psi_n(z)&=& \sqrt{\rho(z-n\Delta z)} J_n(S)\label{Eq:imprint4}\\
& & \cdot e^{i\left[\alpha (z-n\Delta z)^2 + n(\phi-\pi/2) + nk(z-n\Delta z) + \phi_{\text{trap},n} + \phi_{\text{tof},n}\right]}\nonumber
\end{eqnarray}

\begin{figure}
\centerline{\scalebox{0.47}{\includegraphics{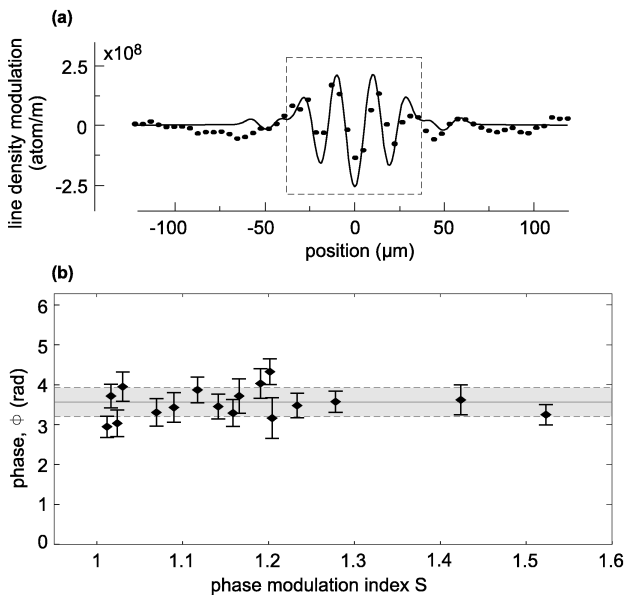}}}
\caption{Interference of overlapping diffraction orders. High
contrast interference fringes are observed for $S=1.1-1.6$. (a)
Line density modulation of the cloud (dots) for S=1.2.
The solid line is the fit by the coherent superposition of the
diffraction orders. (b) The phase $\phi$ is derived
from the fit of the center part of the line density modulation.
Each data point is a single measurement, the error bars indicate
the standard deviation of each fit. The average phase spread is~$\pi/8$.} \label{fig:Figure5}
\end{figure}

We restrict our analysis to the reproducibility of the
interference fringes for different realizations of the experiment.
A reproducible spatial density distribution of atoms is a direct
signature of phase coherent splitting and manipulation of the
condensate \cite{Shin2004,Shin2005}. The reproducibility of the interference fringes relies
on the reproducibility of the spatial position of the condensate
relative to the lattice. A variation of the position would show up
as a variation of $\phi$. Other phase contributions are set by the
experimental conditions and do not change between successive
realizations of the experiment. By fitting the data with the
coherent superposition of the diffraction orders varying only the
single parameter $\phi$, we find that the relative phase of the
diffraction orders is deterministic with an uncertainty of $\pi/8$
on average (standard deviation). A typical interference pattern is
illustrated in (Fig.\ref{fig:Figure5}a).
It shows the line density
modulation of the cloud: $\rho_{\text{coh}}-\rho_{\text{incoh}}$.
The phase $\phi$, derived from the central part of the
interference pattern, is plotted in Fig.\ref{fig:Figure5}b. Taking
into account the condensate's axial size of ($\sim 50\mu\text{m}$)
and the number of ``illuminated'' lattice sites ($\sim 12$) the
observed phase spread is "diffraction limited", i.e. given by the
resolution of the lattice $\Delta k/k=1/n$. This uncertainty due
to the finite size of the condensate could be further reduced by
expanding the axial size of the cloud.

From the highly reproducible interference fringes we conclude that
splitting of condensates with an integrated magnetic grating is
phase coherent. In addition, we have proven that the experimental
parameters can be controlled with an accuracy sufficient for the
construction of integrated matter wave interferometers. In
contrast to other schemes, the interferometer presented here
requires a single diffraction event only. The interferometric path
is closed by the ballistic expansion due to mean-field repulsion
of the condensate atoms. For above experimental conditions, the resolution of a force detector is on the order of $10^{-4}\text{g}$. 
The simultaneous interference of more than two diffraction orders allows not only for the detection of forces but also 
force gradients, and even higher spatial derivatives of external
forces.

We thank C. J. Vale, T. Judd, and M. Fromhold for helpful comments on the manuscript.
Financial support from the Landesstiftung Baden-W\"{u}rttemberg,
and the European Union (MRTN-CT-2003-505032) is greatly
acknowledged. 

%\bibliographystyle{apsrev}
%\bibliography{D:/Dissertation/Andreas}

\end{document}